# A Comprehensive JWST/NIRSpec Census of Broad-line Active Galactic Nuclei: Faint, Low-mass Accretors and a Population of Little Red Dots

Caroline J. Baccus[1,2,5] and Xinfeng Xu[3,4]
[1] New York University, 726 Broadway Rm. 1005, New York, NY 10003, USA; cjb9346@nyu.edu
[2] Menlo School, Atherton, CA, USA
[3] Center for Interdisciplinary Exploration & Research in Astrophysics (CIERA), Evanston, IL, USA
[4] Department of Physics and Astronomy, Northwestern University, 2145 Sheridan Road, Evanston, IL, 60208, USA
*Received 2025 December 2; revised 2026 June 4; accepted 2026 June 7; published 2026 July 22*

## Abstract

We present a sample of 227 broad-line active galactic nuclei (BLAGNs), incorporating 148 newly identified sources, spanning a redshift interval from $z = 0.8$ to 7.2. We analyze spectroscopic data from the NIRSpec instrument on board the James Webb Space Telescope, using the G140H, G140M, G235H, G235M, G395H, and G395M gratings to survey $N \sim 80{,}000$ galaxies for BLAGNs. Through emission-line fitting, using a sum of Gaussian models for H$\alpha$, H$\beta$, [N II] $\lambda\lambda$6548, 6584, and [O III] $\lambda\lambda$4959, 5007, we separate active galactic nucleus (AGN) broad-line components from narrow-line emission. We find the detection rate of BLAGNs to be relatively consistent across our redshift range. Compared to typical low-$z$ AGNs ($z \lesssim 1$), high-$z$ BLAGNs are systematically fainter and less massive, yet they accrete more efficiently, with most showing Eddington ratios between 0.1 and 1.0. This confirms rapid black hole growth during the early cosmic epochs. The detection of faint, low-mass BLAGNs at high redshift also helps bridge the observational gap between local supermassive black holes and remote luminous quasars, providing a more complete view of black hole–galaxy coevolution across cosmic time. We also identify 48 known little red dots (LRDs) in our BLAGN sample. For these LRDs, we find a systematic preference for exponential over Gaussian broad-line profiles, which are consistent with electron scattering rather than virial broadening. This implies that standard virial black hole masses for LRDs may be overestimated by up to 2 orders of magnitude.

## 1. Introduction

Active galactic nuclei (AGNs) powered by supermassive black holes (SMBHs) are among the most luminous and energetic phenomena in the cosmos. They drive galaxy growth and shape evolutionary pathways through intense radiation, jets, and outflows (e.g., J. Silk & M. J. Rees 1998; M. J. Hardcastle & J. H. Croston 2020; S. Laha et al. 2021). Observations and theoretical studies show that SMBHs coevolve with their host galaxies, with black hole accretion and stellar-mass assembly regulating each other (e.g., J. Kormendy & L. C. Ho 2013; N. J. McConnell & C.-P. Ma 2013). Shaped by the interplay of dark matter halos, gravitational forces, and feedback from supernovae and AGNs, their coevolution represents a fundamental process governing galaxy evolution across cosmic time (e.g., P. F. Hopkins et al. 2006; A. C. Fabian 2012; G. Mountrichas 2023).

Despite extensive observational efforts to study AGNs across a wide range of redshifts (e.g., N. A. Grogin et al. 2011; I. Pâris et al. 2018; E. Chaussidon et al. 2023), analyses of high-redshift AGNs ($z \gtrsim 4$) before the launch of James Webb Space Telescope (JWST) are limited by various observational challenges. At high redshift, key AGN diagnostic lines such as [O III] $\lambda\lambda$ 4959, 5007, and H$\alpha$ are redshifted into the near-infrared (NIR), where ground-based spectroscopy suffers strong atmospheric absorption and telluric emission (e.g., P. Rousselot et al. 2000; R. I. Davies 2007). In addition, faintness of high-redshift AGNs in the NIR limits observations to the most luminous sources (Y. Matsuoka et al. 2019; M. Onoue et al. 2019; Y. Shen et al. 2019). This luminosity-driven selection biases samples toward high-accretion-rate SMBHs and may not represent the full AGN population at high redshift (e.g., D. J. Mortlock et al. 2011; E. Banados et al. 2018).

Space-based observatories, such as the Hubble Space Telescope (HST), made the first significant breakthroughs by avoiding atmospheric interference and probing the Universe within the first few hundred million years after the Big Bang (e.g., S. V. W. Beckwith et al. 2006). For example, HST's ultradeep surveys first uncovered GN-z11 at $z$~11, establishing the pre-JWST record for the most distant galaxy known (R. S. Ellis et al. 2013; P. A. Oesch et al. 2016; A. J. Bunker et al. 2023). However, with its 2.4 m mirror and infrared cutoff ~2.5 $\mu$m, HST's sensitivity to faint, rest-frame optical emission of high-redshift AGNs remains limited, leaving the faint AGN population largely unexplored (but see, e.g., A. M. Koekemoer et al. 2011; R. A. Windhorst et al. 2011).

Though launched less than 5 yr ago, JWST has already revolutionized our view of early Universe AGNs. Compared to previous infrared space telescopes, such as Spitzer, JWST's 6.5 m primary mirror—approximately 60× Spitzer's collecting

[5] Corresponding author.

area—provides greatly enhanced infrared sensitivity. Additionally, JWST's extended wavelength coverage up to 30 $\mu$m enables it to detect high-redshift AGNs more effectively than HST. These advancements allow JWST to reveal substantial population of lower-luminosity AGNs at $z \gtrsim 3$–4, pushing into regimes previously inaccessible (e.g., D. D. Kocevski et al. 2023b; Y. Harikane et al. 2023; D. D. Kocevski et al. 2023a; J. Matthee et al. 2024; W. He et al. 2024; R. Maiolino et al. 2024; A. J. Taylor et al. 2025; I. Juodžbalis et al. 2026). Specifically, programs including JWST Advanced Deep Extragalactic Survey (JADES) and Cosmic Evolution Early Release Science (CEERS) conduct deep NIRSpec spectroscopy that identifies faint AGNs with masses of $10^5$–$10^8 M_\odot$, filling the gap between local SMBHs and the luminous quasars seen in previous high-redshift studies (e.g., D. D. Kocevski et al. 2023b; Y. Harikane et al. 2023; D. D. Kocevski et al. 2023a; R. Maiolino et al. 2024). Together, these observations highlight JWST's capability to detect larger populations of AGNs and constrain their properties at high redshift, offering unprecedented insight into AGN evolution in the early Universe.

In addition, recent findings from JWST have revealed that compared to local AGNs, high-$z$ AGNs exhibit higher bolometric luminosities and Eddington ratios as a function of black hole mass, suggesting higher accretion efficiencies and rapid black hole growth (e.g., Y. Harikane et al. 2023; H. Übler et al. 2023; E. Lambrides et al. 2026; H. Suh et al. 2025). These results suggest that faint AGNs may represent an early evolutionary phase in black hole growth and that SMBHs in the early Universe grow faster than their host galaxies, enriching the Universe with higher metallicity (e.g., H. Übler et al. 2023). However, the majority of these conclusions are based on single or small sample high-redshift AGNs (e.g., H. Suh et al. 2025). Therefore, their conclusions could be influenced by selection bias as AGNs with higher accretion rates are intrinsically brighter and more easily detected, potentially leading to an overgeneralization of the accretion rates to all high-redshift AGNs.

In this work, we perform such a systematic search of broad-line active galactic nuclei (BLAGNs) using the v4 reduction of the Dawn JWST Archive (DJA).[6] We describe the observations and the DJA database in Section 2. We then present the sample selection, emission-line fitting, and dust extinction correction in Section 3. After that, we present the main results in Section 4, including the catalog of identified BLAGNs and key measurable properties, including line kinematics, BH mass, luminosity, and Eddington ratio. Finally, we discuss how the properties of BLAGNs evolve over redshift in Section 5. We conclude the paper in Section 6.

We adopt a flat $\Lambda$CDM cosmology with $H_0 = 70$ km s$^{-1}$ Mpc$^{-1}$ and $\Omega_M = 0.3$, according to Planck Collaboration et al. (2020).

## 2. Observations and Data

DJA is a publicly available repository of JWST, including reduced data and products, maintained by Gabriel Brammer. We adopt their most recent data release (G. Brammer & F. Valentino 2025; F. Valentino et al. 2025).[7] They reduce the raw JWST data using a combination of the standard JWST Science Calibration Pipeline and custom reduction routines implemented in the `msaexp` package, the details of which are described in K. E. Heintz et al. (2024) and A. de Graaff et al. (2025).

In brief, raw exposures were bias-subtracted and flat-fielded, then corrected for vignetting and both local and global background. Each slit spectrum was centered, sampled onto a uniform wavelength grid, and extracted from 2D into 1D. Flux-loss corrections for extended profiles were applied via a simple source-model prescription.

From the full DJA catalog of ∼ 80,000 galaxies, we selected all targets observed with NIRSpec's high- (G140H, G235H, G395H) and medium-resolution (G140M, G235M, G395M) gratings for medium-to-high-redshift ($z \gtrsim 2$) targets. We then apply redshift-reliability cuts (i.e., grade >2.5 from DJA) to exclude extractions with ambiguous redshift solutions. The selected galaxies serve as the parent sample for our emission-line fitting in Section 3. These galaxies are drawn primarily the following JWST programs: AURORA (A. E. Shapley et al. 2025), BLUEJAY (S. Belli et al. 2024), CANUCS (G. Rihtaršič et al. 2025), CECILIA (A. L. Strom et al. 2023), CEERS (S. L. Finkelstein et al. 2025), EXCELS (A. C. Carnall et al. 2024), GLIMPSE (V. Kokorev et al. 2025), JADES (D. J. Eisenstein et al. 2026), LyC22 (D. Schaerer et al. 2021), RUBIES (A. de Graaff et al. 2025), UNCOVER (J. R. Weaver et al. 2024), and VALENTINO (K. Ito et al. 2026).

## 3. Analysis

### 3.1. Selection Criteria

To select BLAGNs, we require objects to satisfy the following criteria (see, e.g., Y. Harikane et al. 2023):

1. The full width at half-maximum of the broad component ($\mathrm{FWHM_{broad}}$) of H$\alpha$ (if unavailable, H$\beta$) exceeds 1000 km s$^{-1}$.
2. The signal-to-noise ratio (SNR) of the broad component of H$\alpha$ (if unavailable, H$\beta$) exceeds 5.
3. $\mathrm{FWHM_{broad}}$ of the forbidden lines ([O III] and [N II]) is below 500 km s$^{-1}$.
4. The inclusion of a broad component is statistically justified by an F-test (see Section 3.2).

The threshold of $\mathrm{FWHM_{broad}} > 1000$ km s$^{-1}$ is generally adopted to ensure high specificity to AGN-driven broad-line regions, which are not observed in purely star-forming galaxies (e.g., Y. Harikane et al. 2023).

### 3.2. Emission-line Fitting

To search for BLAGN signatures, we fit Gaussian profiles to strong emission lines (Y. Harikane et al. 2023; A. J. Taylor et al. 2025). In our analysis, we simultaneously fit the H$\alpha$ + [N II]$\lambda\lambda$ 6548, 6584 complex, then the H$\beta$ + [O III] $\lambda\lambda$ 4959, 5007 complex. The local continuum is determined by a linear baseline fit to sideband intervals bracketing each complex and subtracted prior to fitting the emission lines.

We first perform a single-Gaussian fit (hereafter, model–1) to each line to serve as a baseline for the F-test. Second, we fit four Gaussian profiles to each line complex (hereafter, model–2). For the primary lines, i.e., H$\alpha$ and H$\beta$, we fit them with a double Gaussian profile, including one broad and one narrow component. For other lines, i.e., [N II] and [O III], we fit them with a single-Gaussian profile. For the latter, a single Gaussian is sufficient as the high-density gas in BLAGNs suppresses the broad-component emission from these forbidden lines. Overall, the four-Gaussian-profile fitting follows the function:

$$f(\lambda) = \sum_i A_i \frac{1}{\sigma_i \sqrt{2\pi}} \exp\left(-\frac{(\lambda - \lambda_{i0} - \lambda_i)^2}{2\,\sigma_i^2}\right), \quad (1)$$

where $\lambda_{i0}$ is the expected central wavelength for the primary line, $\lambda_i$ is the center shift of other lines in respect to the primary line, $A_i$ is the amplitude of each line, and $\sigma_i$ is the velocity dispersion. An example of the fitting is shown in Figure 1, where we show the data in black and error in gray. The green and blue lines represent the narrow and broad components, respectively, while the red lines are for the total flux combining the two components.

[6] https://dawn-cph.github.io/dja/spectroscopy/nirspec/

[7] https://zenodo.org/records/15472354

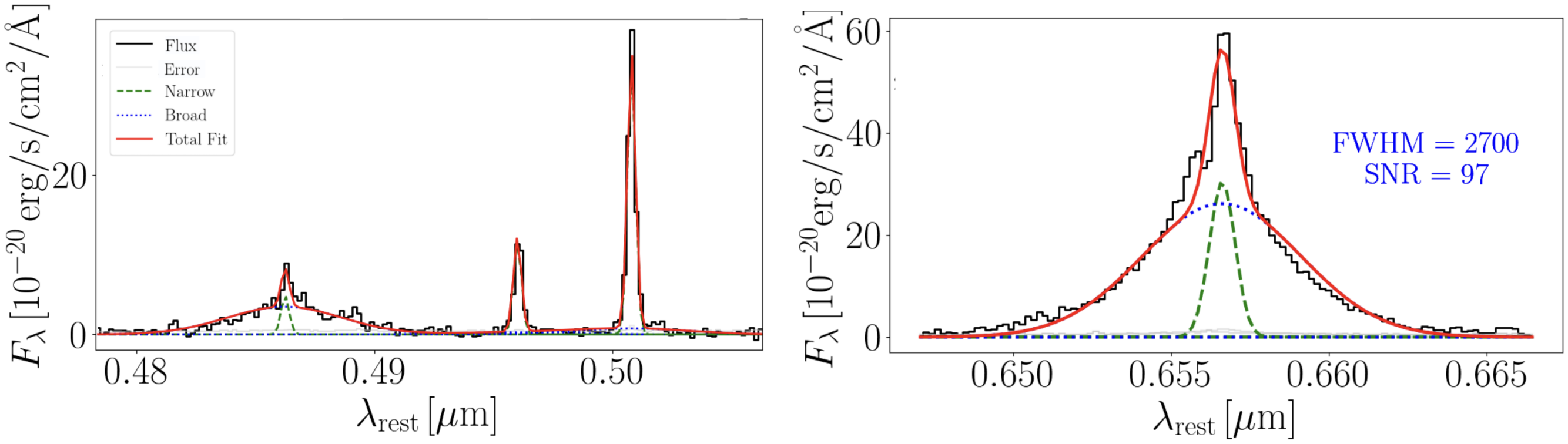

**Figure 1.** Gaussian fittings to detect BLAGN signatures for VALENTINO_3567_51909 ($z = 4.626$). We show the fittings for [O III] $\lambda\lambda$ 4959, 5007 + H$\beta$ lines (left), and H$\alpha$ + [N II]$\lambda\lambda$ 6548, 6584 (right). The NIRSpec data and errors are shown as black and gray lines, respectively. The green and blue lines are for the narrow and broad component, respectively. The red lines are the sum of both components. We also show the SNR and FWHM of the broad H$\alpha$ component in the right panel. See Section 3.2 for the fitting details.

To assess the significance of the broad components, we perform an F-test using the `scipy.stats.f.ppf` function as follows:

$$F = \frac{\chi_1^2 - \chi_2^2}{p_2 - p_1} \Big/ \frac{\chi_2^2}{n - p_2}, \quad (2)$$

where $\chi_1^2$ and $\chi_2^2$ are the sum of squared residuals for model–1 and model–2, respectively. The change in degrees of freedom is given by $p_2 - p_1$. $n$ is the number of data points used in the fits. We adopt a significance level $\alpha = 0.05$ and reject the null hypothesis (that model–1 suffices) whenever the measured $F$ exceeds the critical value $F_\alpha(p_2 - p_1,\ n - p_2)$ from the $F$-distribution.

We fix the flux ratio for doublet lines, e.g., [N II] 6584/6548 = 2.93 and [O III] 5007/4959 = 2.98, according to atomic physics (e.g., P. J. Storey & C. J. Zeippen 2000; D. E. Osterbrock & G. J. Ferland 2006). Model parameters are optimized via maximum-likelihood fitting using the error spectrum, testing Powell, L-BFGS-B, and Nelder–Mead algorithms and adopting the solution that minimizes $\chi^2$.

For the medium-resolution data, we also correct the measured broad-line FWHM for instrumental broadening by subtracting the NIRSpec line-spread function (LSF) in quadrature:

$$\mathrm{FWHM_{int}} = \sqrt{\mathrm{FWHM_{mea}^2} - \mathrm{FWHM_{LSF}^2}}, \quad (3)$$

where $\mathrm{FWHM_{mea}}$ is the measured FWHM, $\mathrm{FWHM_{LSF}}$ is the instrumental FWHM and is linearly interpolated at the observed wavelength of H$\alpha$ using the dispersion curves from the JWST User Documentation,[8] and $\mathrm{FWHM_{int}}$ is the final corrected FWHM. We adopt a Monte Carlo approach to estimate uncertainties on all model parameters. We perform $N = 1000$ simulations, in each of which the observed fluxes are perturbed by Gaussian noise with zero mean and standard deviation equal to the $1\sigma$ flux error. Each perturbed spectrum is then refit with our emission-line model. The $1\sigma$ uncertainties on each parameter are taken as the 16th and 84th percentiles of the resulting parameter distributions.

We note that some of our objects have both medium- and high-resolution spectra. By comparing them, we find the higher resolution provides access to kinematic information that is unresolved in medium-resolution data. For example, it enables robust constraints on the line profiles, more accurate black hole mass estimates from line widths, and the detection of weak emission lines (e.g., H$\beta$, [N II]) that are blended or undetectable at medium resolution. Therefore, we adopt the high-resolution spectra whenever available.

### *3.3. Dust Extinction Correction*

To correct the observed emission-line fluxes for dust attenuation, we compute the color excess $E(B - V)$ using the Balmer decrement, defined as the ratio of the integrated flux of H$\alpha$ to H$\beta$ (broad + narrow components). Since most of our sources do not have a high SNR Hb line, the two-component decomposition is usually not possible for H$\beta$. Thus, the total flux is adopted in the calculations. This also provides a model-independent dust measurement, avoiding the systematic uncertainties associated with decomposing the line into narrow and broad components. Note that the broad- and narrow-line regions of an AGN may have different dust contents along their lines of sight (M. Brooks et al. 2025), so the true correction for the broad component may differ from that inferred from the total line ratio. The luminosities reported in Section 4.3 should be interpreted with this caveat.

We adopt an intrinsic ratio of 3.1, which is more appropriate for the gas conditions in both the broad-line and narrow-line regions of AGNs (see D. E. Osterbrock & G. J. Ferland 2006; M. Brooks et al. 2025). The color excess is defined following

[8] https://jwst-docs.stsci.edu/

D. Calzetti et al. (2000) as

$$E(B - V) = 1.97 \log_{10}\left(\frac{F_{H\alpha}/F_{H\beta}}{3.1}\right), \quad (4)$$

where $F_{H\alpha}$ and $F_{H\beta}$ are the measured fluxes of components of the H$\alpha$ and H$\beta$ emission lines (M. Brooks et al. 2025). Using the derived $E(B - V)$, we compute the visual extinction $A_V$ assuming an extinction ratio of $R_V = 3.1$ (D. E. Osterbrock & G. J. Ferland 2006) by

$$A_V = R_V \cdot E(B - V). \quad (5)$$

We then apply the D. Calzetti et al. (2000) attenuation curve to derive the wavelength-dependent extinction for the whole spectrum:

$$A_\lambda = k(\lambda) \cdot E(B - V), \quad (6)$$

where $k(\lambda)$ is the Calzetti attenuation law evaluated at rest-frame wavelengths. The extinction is applied to deredden the fluxes of all emission lines using the the Python package `extinction`.

For objects whose spectra do not cover H$\beta$, we adopt the median $E(B - V) \sim 0.585$ of the rest of the sample.

## 4. Results

### 4.1. Identified BLAGNs

Based on these selection criteria, we identify 227 BLAGNs. These objects are listed in Table 1 and their Gaussian fitting results are shown in Figure 8 in the Appendix. Among the identified BLAGNs, 79 sources have been previously reported in the literature. Within our sample, 13 AGNs are identified from the G395H grating, 14 from G235H, 76 from G235M, 110 from the G395M grating, and 14 from the G140M grating. All identified BLAGNs exhibit prominent broad-line emission in the H$\alpha$ line or, where available, the H$\beta$ line.

In Figure 2, we show the redshift distribution of our identified BLAGNs (blue histogram), where we overlay the histograms for our parent sample from DJA v4 database in red. Comparing the two distributions, we find the BLAGN detection rate is approximately uniform between $z = 2$ and 7, with the highest-redshift detections at $z \sim 7.2$. The decline at $z > 7$ could be because H$\alpha$ is redshifted to the edge of NIRSpec's G395 grating, so the identification becomes more challenging.

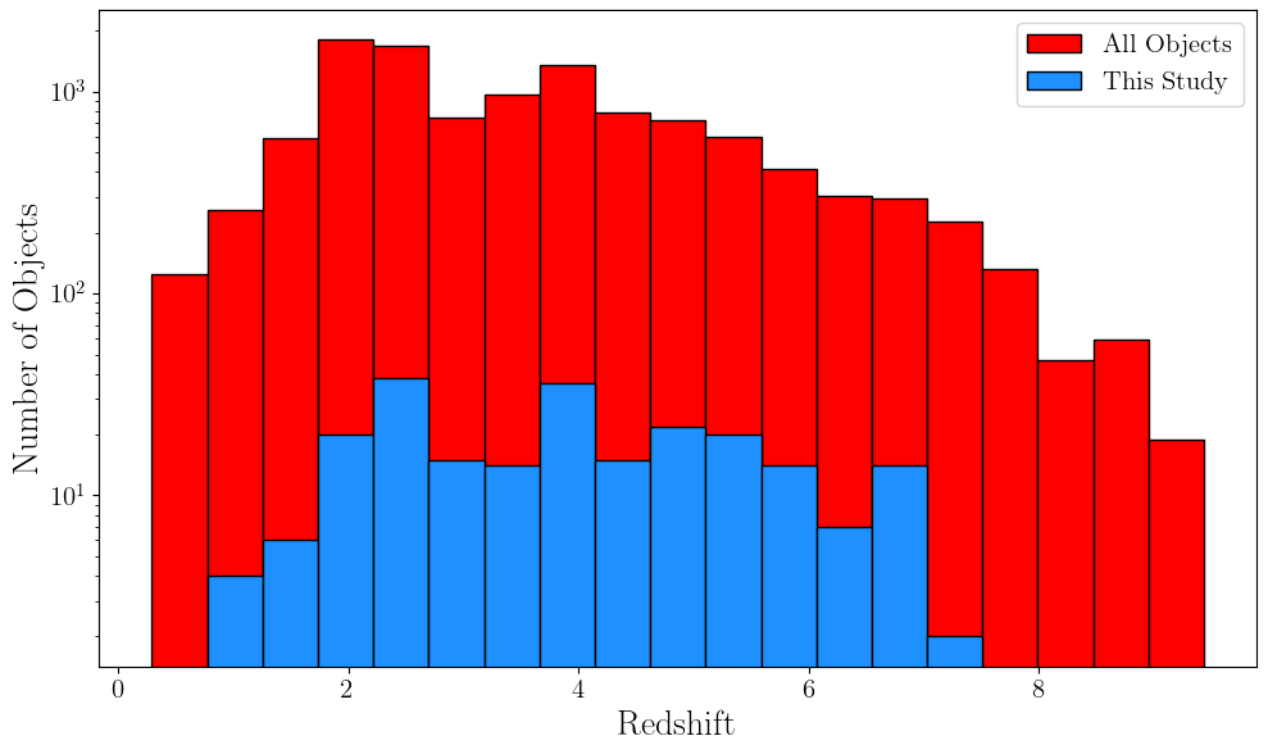

**Figure 2.** The redshift distribution of our sample of BLAGNs (blue) and the total objects searched in the DJA v4 database (red) on a log scale. The detection rate of BLAGNs is almost uniform between $z = 2$ and 7, with the highest-redshift objects out to $z = 7.2$.

### 4.2. BH Mass–Luminosity Relationship

For each identified BLAGN, we then derive key physical parameters such as the BH mass ($M_{BH}$) and bolometric luminosity ($L_{bol}$) as follows.

We first computed the luminosity distance $D_L$ for each galaxy, which can be adopted to calculate the H$\alpha$ luminosity as

$$L_{H\alpha} = F_{H\alpha} \times 4\pi D_L^2, \quad (7)$$

where $F_{H\alpha}$ is the total broad H$\alpha$ line flux, after correcting by the $E(B - V)$ values for each object as detailed in Section 3.3. If H$\alpha$ are not covered, we adopt $F_{H\alpha} = F_{H\beta} \times 3.1$ instead.

We then estimate $M_{BH}$ using the virial method calibrated at $z \sim 0$ (J. E. Greene & L. C. Ho 2005; J. Matthee et al. 2024), which yields

$$M_{BH} = 2.0 \times 10^6 \left(\frac{L_{H\alpha}}{10^{42}\ \mathrm{erg\ s^{-1}}}\right)^{0.55} \left(\frac{\mathrm{FWHM}_{H\alpha}}{10^3\ \mathrm{km\ s^{-1}}}\right)^{2.06} M_\odot, \quad (8)$$

where $L_{H\alpha}$ and $\mathrm{FWHM}_{H\alpha}$ are the luminosity and FWHM of the broad component of the H$\alpha$ emission line. It is important to note that $\sim$0.5 dex systematic uncertainty applies to this relation at high redshifts (A. E. Reines & M. Volonteri 2015; R. Abuter et al. 2024).

We then estimate $L_{bol}$, i.e., the total energy output of the AGNs, by first scaling the $L_{H\alpha}$ to the 5100 Å continuum using empirical relations, with a bolometric correction of 9.8 for standard AGNs (R. J. McLure & J. S. Dunlop 2004; G. Risaliti & M. Elvis 2004; Y. Harikane et al. 2023; K.-X. Lu et al. 2025) and a bolometric correction of 5.4 according to J. E. Greene et al. (2026) for objects previously identified as little red dots (LRDs) in the literature (see Section 5.3 for more discussion):

$$L_{bol} = 9.8 \times L_{5100}. \quad (9)$$

Figure 3 shows the $L_{bol}$ and BH mass of our sample of AGNs (with star symbols) and ones from the literature. Our AGNs have bolometric luminosities ranging from $10^{43}$ to $10^{45.5}$ erg s$^{-1}$ and BH masses ranging from $10^5$ to $10^9\ M_\odot$, which are much lower on average than lower-redshift AGNs. The latter are shown as the contours from Sloan Digital Sky Survey (SDSS) Data Release 16 (DR16; Q. Wu & Y. Shen 2022), where the three contours represent the density distribution of AGNs enclosing 68%, 95%, and 99.7% of the total sample. This difference suggests that we cover much smaller and fainter BLAGNs in JWST. This is consistent with the findings in previous studies of high-redshift BLAGNs using deep JWST observations (e.g., Y. Harikane et al. 2023; R. Maiolino et al. 2024), but with a substantially larger sample of 227 objects compared to theirs (10 and 71, respectively).

In addition, there is a strong positive correlation between BH mass and $L_{bol}$, which is consistent with expectations from previous studies and models of BH accretion (e.g., Y. Harikane et al. 2023; R. Maiolino et al. 2024; A. J. Taylor et al. 2025). We also plot lines showing Eddington ratios ($\lambda_{edd}$) and find that our identified BLAGNs mostly have $\lambda_{edd}$ between 0.1 and 1.0, indicating that they are highly accreting.

**Table 1**
Broad-line AGN Properties in Our Sample

| Object ID | R.A. | Decl. | $z$ | $\mathrm{FWHM}_{\mathrm{H}\alpha,\mathrm{broad}}$ (km $s^{-1}$) | $\mathrm{Log}(M_{\mathrm{BH}})$ ($M_{\odot}$) | $\mathrm{Log}(L_{\mathrm{bol}})$ (erg $s^{-1}$) | $\lambda_{\mathrm{edd}}$ | Grating | LRD | References |
|---|---|---|---|---|---|---|---|---|---|---|
| BLUEJAY_1810_12020 | 150.074554 | 2.301999 | 2.0934 | $4076^{+6}_{-6}$ | $7.82^{+0.0029}_{-0.0035}$ | $44.8^{+0.0043}_{-0.0042}$ | $0.0740^{+0.0003}_{-0.0002}$ | G235M | ⋯ | 1–5 |
| BLUEJAY_1810_11337 | 150.119519 | 2.295778 | 2.0950 | $2640^{+25}_{-23}$ | $7.32^{+0.0017}_{-0.0016}$ | $44.7^{+0.0026}_{-0.0026}$ | $0.1713^{+0.0006}_{-0.0005}$ | G140M | ⋯ | 1,2, 6,7 |
| GTO_1214_1158 | 150.132576 | 2.212071 | 2.1083 | $3736^{+383}_{-877}$ | $7.49^{+0.039}_{-0.0443}$ | $44.5^{+0.054}_{-0.0604}$ | $0.0720^{+0.0028}_{-0.0034}$ | G235H | ⋯ | 8 |
| GDS_1207_209962 | 53.131145 | −27.773187 | 2.2245 | $2140^{+24}_{-27}$ | $7.18^{+0.0024}_{-0.0023}$ | $44.7^{+0.0038}_{-0.004}$ | $0.2799^{+0.0014}_{-0.0014}$ | G235M | ⋯ | 9–14 |
| JADES_1181_28074 | 189.064589 | 62.273815 | 2.2600 | $3836^{+30}_{-30}$ | $7.82^{+0.0028}_{-0.0034}$ | $44.9^{+0.0045}_{-0.0053}$ | $0.0886^{+0.0003}_{-0.0004}$ | G235M | ⋯ | 15–26 |
| LYC22_1869_4096 | 334.251380 | 0.332309 | 2.2762 | $4084^{+1}_{-13}$ | $8.3^{+0.0018}_{-0.0017}$ | $45.5^{+0.0028}_{-0.0026}$ | $0.1348^{+0.0004}_{-0.0003}$ | G235M | ⋯ | 27 |
| BLUEJAY_1810_11142 | 150.073292 | 2.293290 | 2.4465 | $2652^{+175}_{-136}$ | $6.82^{+0.0205}_{-0.0182}$ | $43.8^{+0.033}_{-0.0281}$ | $0.0785^{+0.0021}_{-0.0019}$ | G235M | ⋯ | 3, 5, 28 |
| CEERS_1345_1244 | 215.240652 | 53.036041 | 4.4779 | $2542^{+86}_{-64}$ | $7.21^{+0.0046}_{-0.0063}$ | $44.6^{+0.0063}_{-0.0061}$ | $0.2039^{+0.002}_{-0.0017}$ | G395M | LRD | 22, 29–34 |
| RUBIES_4233_50052 | 214.823454 | 52.830277 | 5.2394 | $2013^{+113}_{-96}$ | $6.82^{+0.0089}_{-0.01}$ | $44.5^{+0.0086}_{-0.0079}$ | $0.3641^{+0.0051}_{-0.0043}$ | G395M | LRD | 29,30, 32, 34–38 |
| GDS_1210_13704 | 53.126535 | −27.818092 | 5.9189 | $2013^{+204}_{-170}$ | $6.55^{+0.0227}_{-0.0208}$ | $43.9^{+0.0199}_{-0.0163}$ | $0.1981^{+0.0047}_{-0.0044}$ | G395M | LRD | 24, 30,31, 35,36, 39 |

**Note.** The list of identified BLAGNs ordered by redshift. The full version of this table is available in machine-readable format in the online journal. A portion is shown here for guidance regarding its form and function. The object ID is in the format of "Survey-name_Program-ID_Object-ID." Several objects have $\mathrm{FWHM}_{\mathrm{H}\alpha,\mathrm{broad}}$ slightly lower than 1000 km $s^{-1}$, but we still keep them since they were identified as BLAGNs in previous studies in the literature (e.g., Y. Harikane et al. 2023; R. Maiolino et al. 2024). Column (9) indicates the main grating of the observations for detecting the strong emission lines. References for the initial JWST observation programs for these sources: AURORA (A. E. Shapley et al. 2025), BLUEJAY (S. Belli et al. 2024), CANUCS (G. Rihtaršič et al. 2025), CECILIA (A. L. Strom et al. 2023), CEERS (S. L. Finkelstein et al. 2025), EXCELS (A. C. Carnall et al. 2024), GLIMPSE (V. Kokorev et al. 2025), JADES (D. J. Eisenstein et al. 2026), LyC22 (D. Schaerer et al. 2021), RUBIES (A. de Graaff et al. 2025), and UNCOVER (J. R. Weaver et al. 2024). The numbers in the References column refer to the following numbered list: (1) F. Stanley et al. (2018); (2) B. Laloux et al. (2023); (3) R. L. Davies et al. (2024); (4) K. Ito et al. (2026); (5) L. Bugiani et al. (2025); (6) S. Mukae et al. (2020); (7) Y. Pang et al. (2025); (8) R. Uematsu et al. (2025); (9) S. Alberts et al. (2020); (10) A. Vayner et al. (2021); (11) J. Lyu et al. (2022); (12) Y. Wang et al. (2024); (13) R. Zhang et al. (2025); (14) G. H. Rieke et al. (2025); (15) I. Juodžbalis et al. (2023); (16) F. Pacucci & R. Narayan (2024); (17) B. Wang et al. (2025); (18) K. Inayoshi & R. Maiolino (2025); (19) B. Trefoloni et al. (2025); (20) R. Tripodi et al. (2025); (21) W. Ishibashi et al. (2025); (22) M. Kokubo & Y. Harikane (2025); (23) I. Delvecchio et al. (2025); (24) R. Maiolino et al. (2025); (25) X. Chen et al. (2026); (26) G. Mazzolari et al. (2026); (27) A. Peca et al. (2023); (28) G. Quadri et al. (2025); (29) Y. Harikane et al. (2023); (30) M. Yue et al. (2024); (31) Y. Xu et al. (2025); (32) A. J. Taylor et al. (2025); (33) G. Mazzolari et al. (2025); (34) V. Rusakov et al. (2026); (35) M. A. Latif et al. (2024); (36) M. Brooks et al. (2025); (37) W. L. Tee et al. (2025); (38) A. Torralba et al. (2026); (39) J. Scholtz et al. (2025).

(This table is available in its entirety in machine-readable form in the online article.)

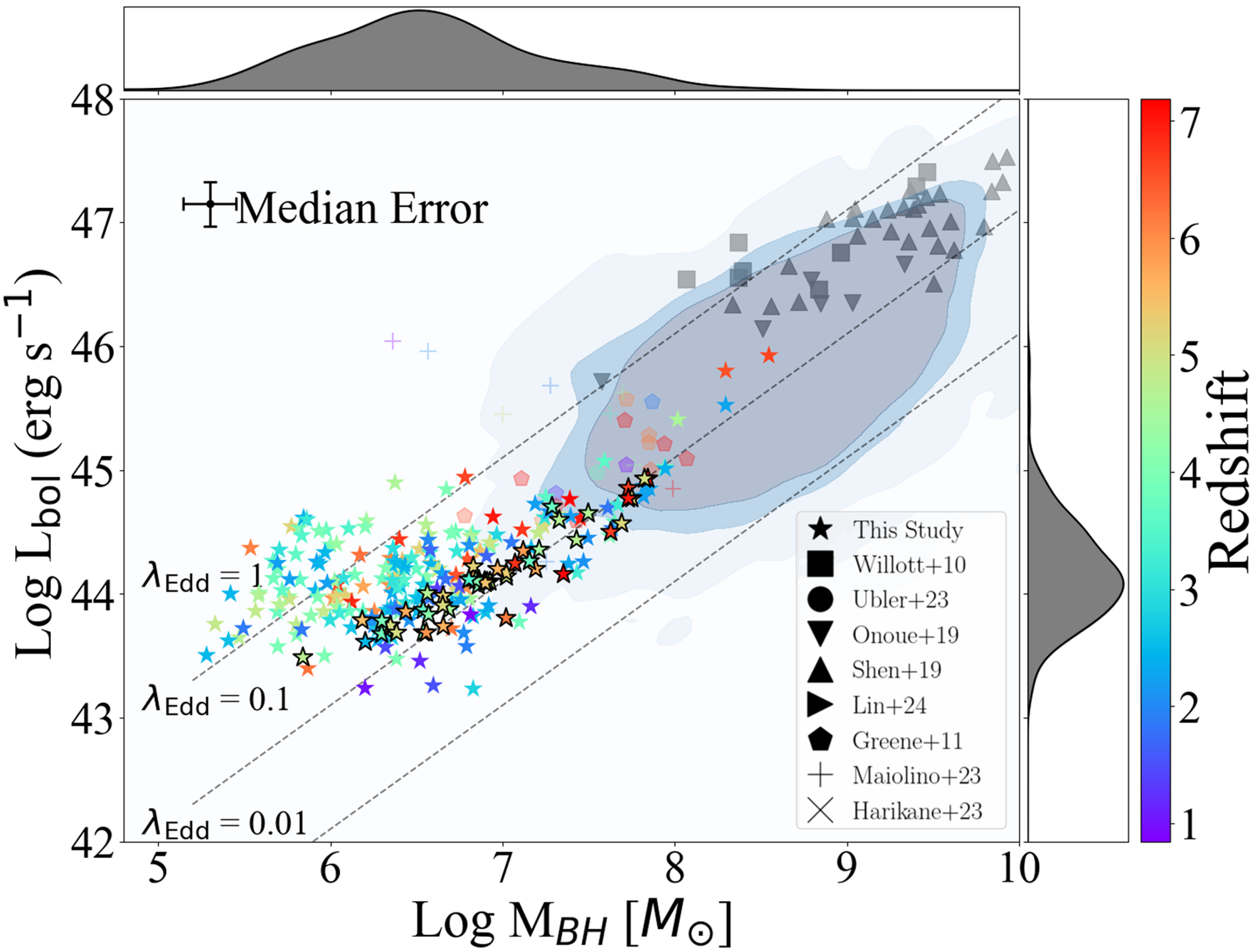

**Figure 3.** Black hole mass ($M_{\rm BH}$) versus bolometric luminosity ($L_{\rm bol}$) for our AGN sample. The color of the points indicates redshift. Symbols indicate AGNs from this study (⋆), C. J. Willott et al. (2010) $z \approx 6$) (■), H. Übler et al. (2023) $z \sim 5.5$) (•), M. Onoue et al. (2019) $6.1 \leqslant z \leqslant 6.7$ (▼), Y. Shen et al. (2019) $z \geqslant 5.7$(▲), X. Lin et al. (2024) $4 \leqslant z \leqslant 5$ (▶), J. E. Greene et al. (2024) $z \geqslant 5$ (⬠), R. Maiolino et al. (2024) $4 \leqslant z \leqslant 11$ (+), Y. Harikane et al. (2023) $4 \leqslant z \leqslant 7$ (×). Light-gray points are studies from which the redshift of the objects is not provided. Contours show low-z AGNs from redshifts 1–2 from SDSS DR16 (Q. Wu & Y. Shen 2022), and the dashed lines show the Eddington ratio ($\square_{\rm edd}$) of 0.01, 0.1, and 1. Points outlined in black are known LRDs from the literature; see Section 5.3.

### 4.3. Velocity Width versus Broad Hα Luminosity

In Figure 4, we show the $\mathrm{FWHM}_{\mathrm{H}\alpha,\mathrm{broad}}$ versus $\log(L_{\mathrm{H}\alpha,\mathrm{broad}})$ for our sample of BLAGNs (star symbols) and others found in the literature. All objects are color-coded by their redshift. We find a weak positive correlation (Spearman's $\rho = 0.351$), indicating that broader H$\alpha$ lines may correspond to higher luminosities. High-$z$ BLAGNs ($z > 4$) also tend to have systematically higher $L_{\mathrm{H}\alpha,\mathrm{broad}}$. To compare with lower-redshift AGNs, we also overlay results from SDSS DR16 in contours. We find that the BLAGNs identified in our sample partially overlap with the low-redshift quasar population, suggesting that the broad-line regions of these high-$z$ BLAGNs may not yet be well distinguished from those of typical Type 1 quasars (see also X. Lin et al. 2024).

### 4.4. Eddington Ratio Evolution with Redshift

Y. Harikane et al. (2023) find a tentative trend that high-$z$ AGNs have higher Eddington ratio ($\lambda_{\rm edd}$). But their sample size is small ($\sim$10) and suggests that more data is needed to confirm this trend. With our larger sample ($\sim$120 AGNs with $z > 4$), we calculate $\lambda_{\rm edd}$ for each AGN similarly as theirs:

$$\lambda_{\rm Edd} = \frac{L_{\rm bol}}{L_{\rm Edd}}, \tag{10}$$

where the Eddington luminosity $L_{\rm Edd}$ is defined as $L_{\rm Edd} = 1.26 \times 10^{38}\,(M_{\rm BH}/M_\odot)\,\mathrm{erg\,s^{-1}}$.

We find there is a clear trend that $\lambda_{\rm edd}$ decreases with increasing $M_{\rm BH}$, indicating that the more massive black holes are accreting less efficiently. This inverse relationship has been tentatively reported in previous high-redshift studies such as Y. Harikane et al. (2023), Y. Aggarwal (2024), and W. He et al. (2024). Combined with the fact that the high-redshift AGNs we identified have much lower $M_{\rm BH}$ than their low-redshift ($z < 1$) counterparts (see contours in Figure 3), this indicates a strong increase in BH accretion rate in the higher-redshift Universe ($z > 2$).

## 5. Discussion

### 5.1. General Properties of BLAGNs over z

Our identification of BLAGNs using JWST NIRSpec spectroscopy significantly expands the known population of AGNs at high redshift (see Figures 2–3) and provides a larger and critical dataset for evaluating BH growth across cosmic times. Previously, similar studies commonly focused on a subset of JWST NIRSpec data (e.g., Y. Harikane et al. 2023; J. E. Greene et al. 2024; R. Maiolino et al. 2024; L. Napolitano et al. 2025) and/or a single survey (e.g., A. D. Goulding et al. 2023; T. C. C. Chien et al. 2024). In contrast, our comprehensive search across the full NIRSpec database enables a more complete census of BLAGNs at high redshift.

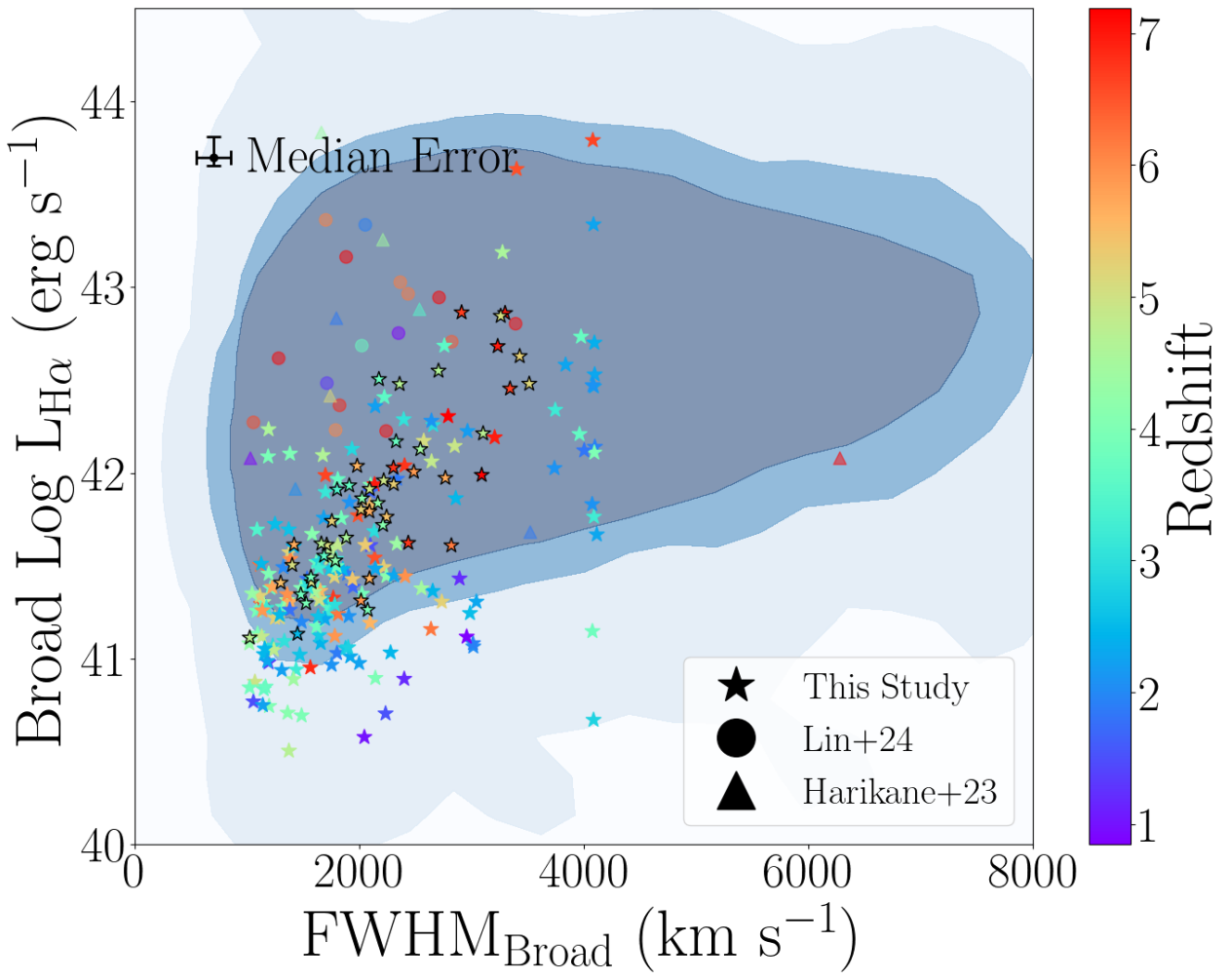

**Figure 4.** $FWHM_{H\alpha,broad}$ versus Log $L_{H\alpha,broad}$ for our sample of BLAGNs. The points are colored by redshift. Along with our BLAGNs, we overlay other BLAGNs found by X. Lin et al. (2024) (•), J. E. Greene et al. (2024) (▲), R. Maiolino et al. (2024) (♦), Y. Harikane et al. (2023) (⬠), B. Trakhtenbrot et al. (2011) (◀) , and C. J. Willott et al. (2010) (▶). BLAGNs identified in this study are shown as ⋆. Similarly as Figure 3, contours show low-z AGNs from redshifts 1–2 (Q. Wu & Y. Shen 2022). Outlined points are known LRDs; see Section 5.3.

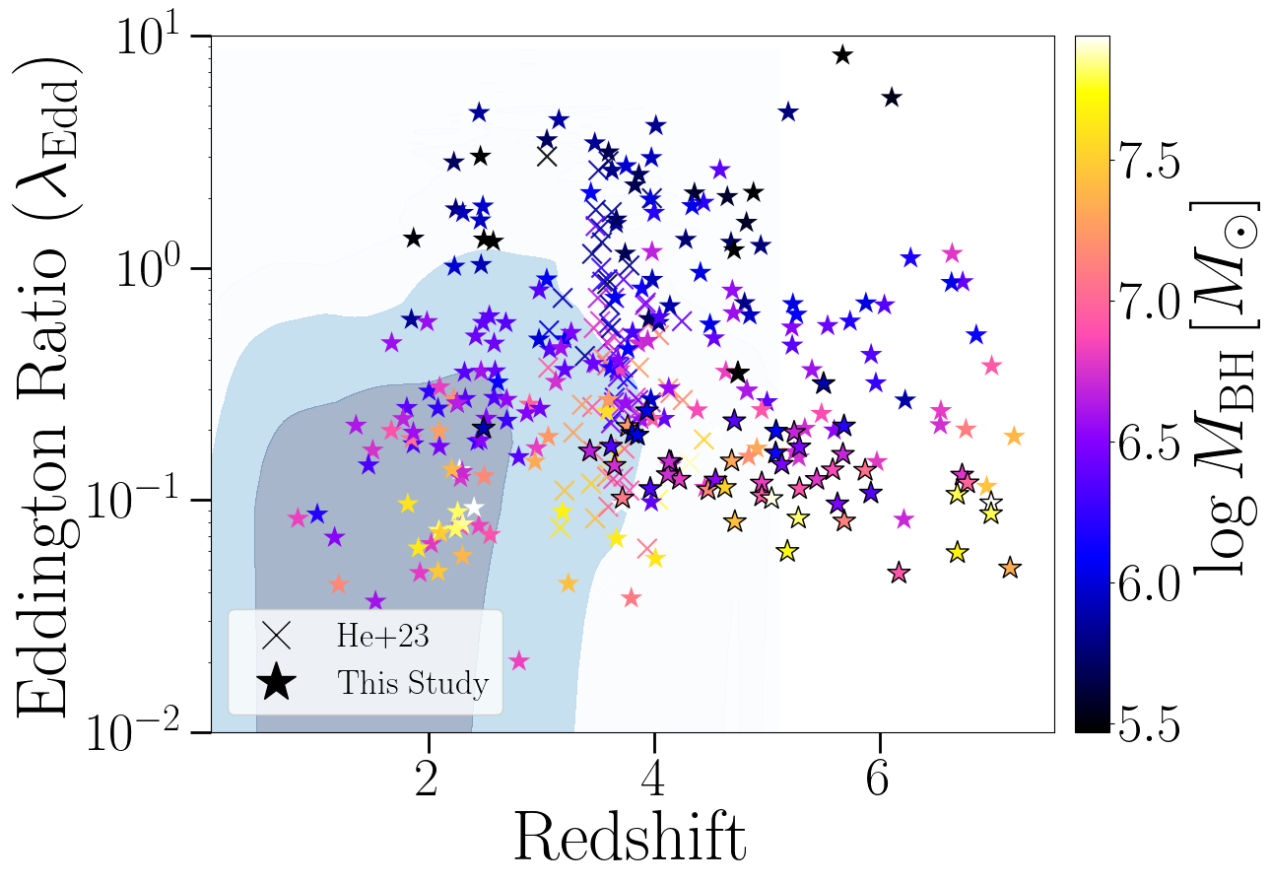

**Figure 5.** Eddington ratio ($\lambda_{edd}$) versus redshift for our sample. The color of the points indicates black hole mass ($M_{BH}$). We also include BLAGNs found in W. He et al. (2024) as crosses. Similarly as Figure 4, we show the lower-redshift AGNs from SDSS DR16 in different contours from Q. Wu & Y. Shen (2022). Points outlined in black are known LRDs from the literature; see Section 5.3.

Specifically, Y. Harikane et al. (2023) identified a smaller sample of higher-redshift AGNs, which suggested that more distant objects exhibited a higher $\lambda_{edd}$ compared to those at $z \sim 0$; however, it was unclear whether this was due to selection bias of brighter AGNs. Our thorough search of the JWST NIRspec archive and our larger BLAGN sample support the conclusion that $\lambda_{edd}$ increases systematically with redshift to the earliest observable time, indicating more rapid black hole growth and accretion in the early Universe (see Figure 5). In addition, we find that high-redshift AGNs are systematically lower in both black hole mass and luminosity than their local counterparts. In Figure 2, we also observe an increasing fraction of BLAGNs out to a redshift of $z \sim 7$. This increasing fraction and elevated Eddington ratios imply that the observed AGN population at $z > 4$ may represent a distinct phase of black hole–galaxy coevolution.

### *5.2. Ionization Mechanisms of the Discovered BLAGNs*

Despite JWST's transformative observational capacity, reliably distinguishing AGN-driven and stellar photoionization in early galaxies remains challenging. The dominant ionization mechanism in a galaxy can be diagnosed using emission-line ratios, particularly [N II]/H$\alpha$ and [O III]/H$\beta$ in a Baldwin-Phillips-Terlevich (BPT) diagram (J. A. Baldwin et al. 1981). However, at high redshift and low metallicity, theoretical models predict that weakened [N II] emission can cause AGN-powered systems to fall within star-forming regions of standard diagnostic diagrams (e.g., H. Übler et al. 2023; F. Arevalo Gonzalez et al. 2025). Consequently, commonly adopted methods such as BPT diagrams may not reliably identify the primary ionization source in early Universe systems, highlighting the need for additional observations. In Figure 6, we show our BLAGNs in the BPT diagram when they have sufficient wavelength coverage to calculate the line ratios. We also overlay two curves that discriminate between galaxies dominated by AGNs and star formation (L. J. Kewley et al. 2001; G. Kauffmann et al. 2003). We find that around half our sources fall into the AGN region in this diagram, while others fall left of the curve of G. Kauffmann et al. (2003). One explanation for the latter is that they have lower metallicity than regular BLAGNs. Future dedicated studies could reveal whether they do have relatively low metallicities.

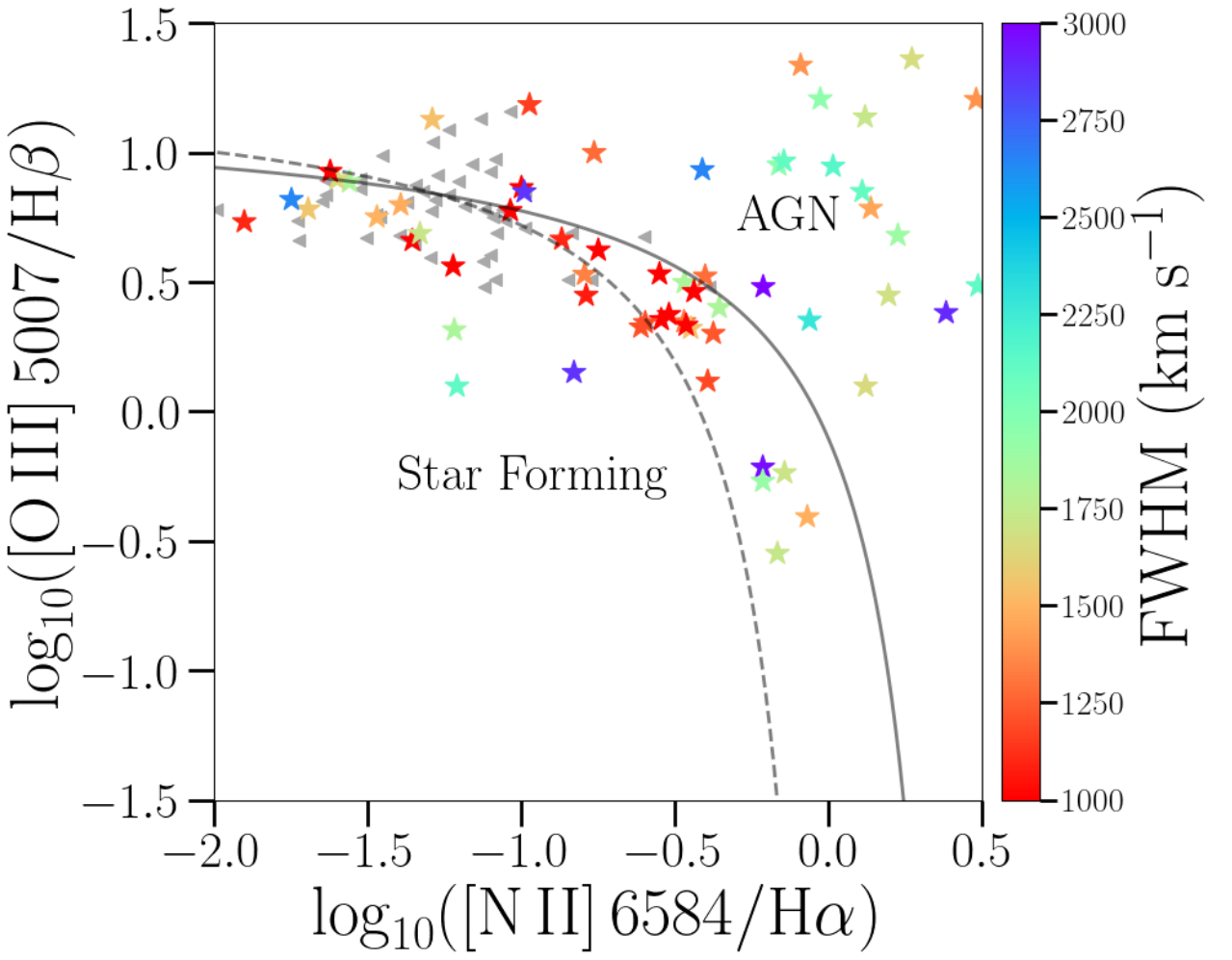

**Figure 6.** BPT diagram of our sample of BLAGNs. ⋆ show points from our study and ▶ represents galaxies $z > 4$ found by K. Nakajima et al. (2023). The black solid and dashed curves show the limits between AGNs and star-forming galaxies as found by L. J. Kewley et al. (2001) and G. Kauffmann et al. (2003).

### *5.3. Properties of LRDs in Our Sample*

LRDs are a recently identified class of compact, red sources at high redshift with unusual properties, such as overmassive BHs and weak X-ray and radio emissions (e.g., Y. Harikane et al. 2023; J. E. Greene et al. 2026; V. Rusakov et al. 2026). We search for LRDs in our sample that have been confirmed in the literature (e.g., D. D. Kocevski et al. 2023a; J. Matthee et al. 2024; H. B. Akins et al. 2025; V. Kokorev et al. 2025; I. Labbe et al. 2025) and find 48 such cases. The physical origin of the broad lines in LRDs remains debated. While bulk Doppler motions from fast-moving gas near an SMBH would produce the characteristic Gaussian line wings expected from virial broadening, electron scattering through a dense, optically thick medium can produce exponential wings with a subtly different profile shape (V. Rusakov et al. 2026). Distinguishing between these two mechanisms is consequential as each implies a fundamentally different black hole mass scale. The virial interpretation can yield overmassive black holes since an electron-scattering interpretation would imply substantially lower intrinsic velocities, and hence smaller masses, by about a factor of 100 (V. Rusakov et al. 2026).

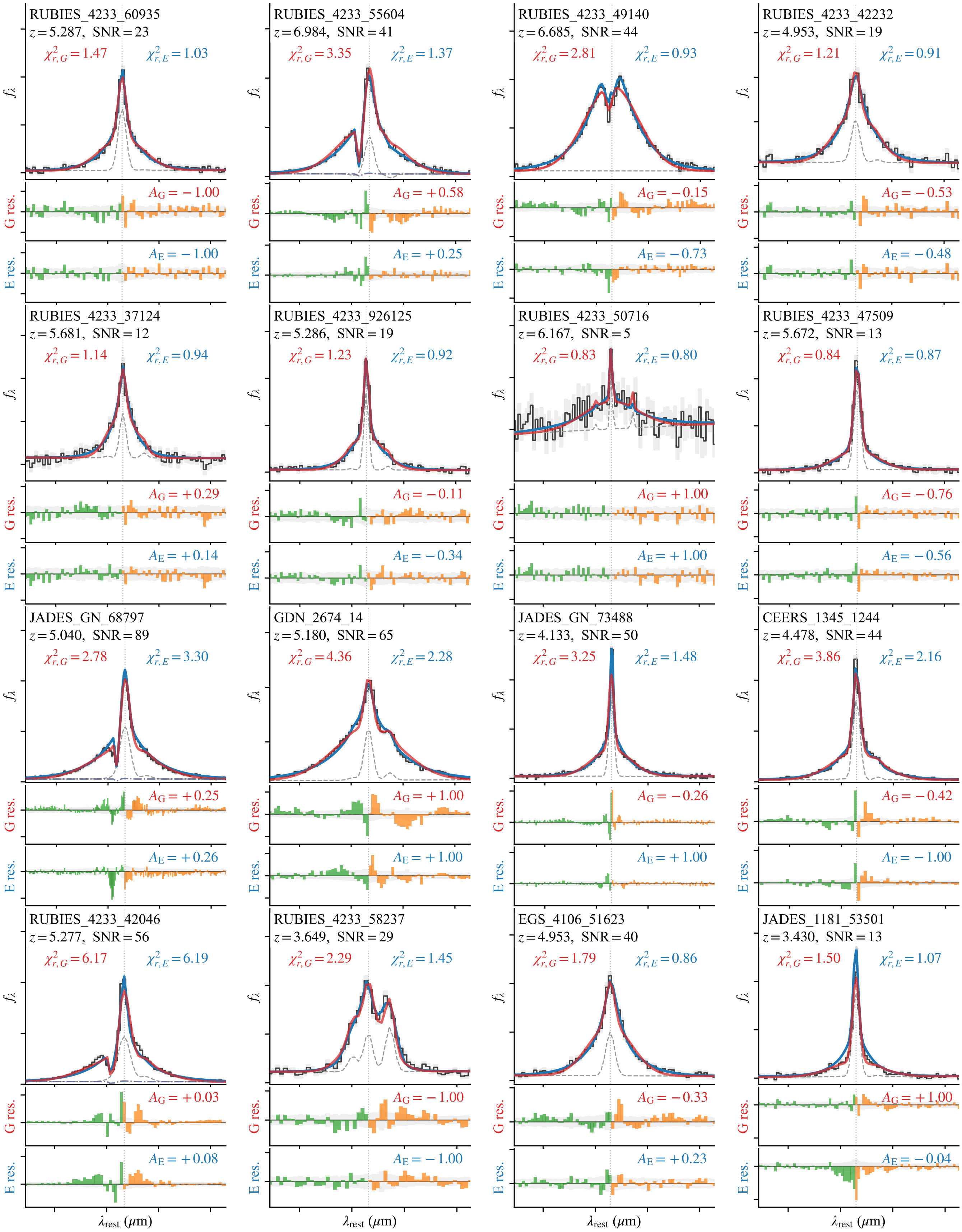

**Figure 7.** Known LRDs from our BLAGN sample. Blue and red lines are the best-fit exponential (V. Rusakov et al. 2026) and Gaussian models, respectively. $\chi^2_{r,G}$ and $\chi^2_{r,E}$ indicate the reduced $\chi^2$ of the Gaussian and exponential fits, respectively. $A_G$ and $A_E$ indicate the broad-line asymmetry, where a value of $-1$ or $+1$ indicates blue- or red-wing excess, respectively. The residuals are shown in the bottom panels, with the left and right wings indicated in green and orange, respectively. Overall, exponential models fit the line profiles better.

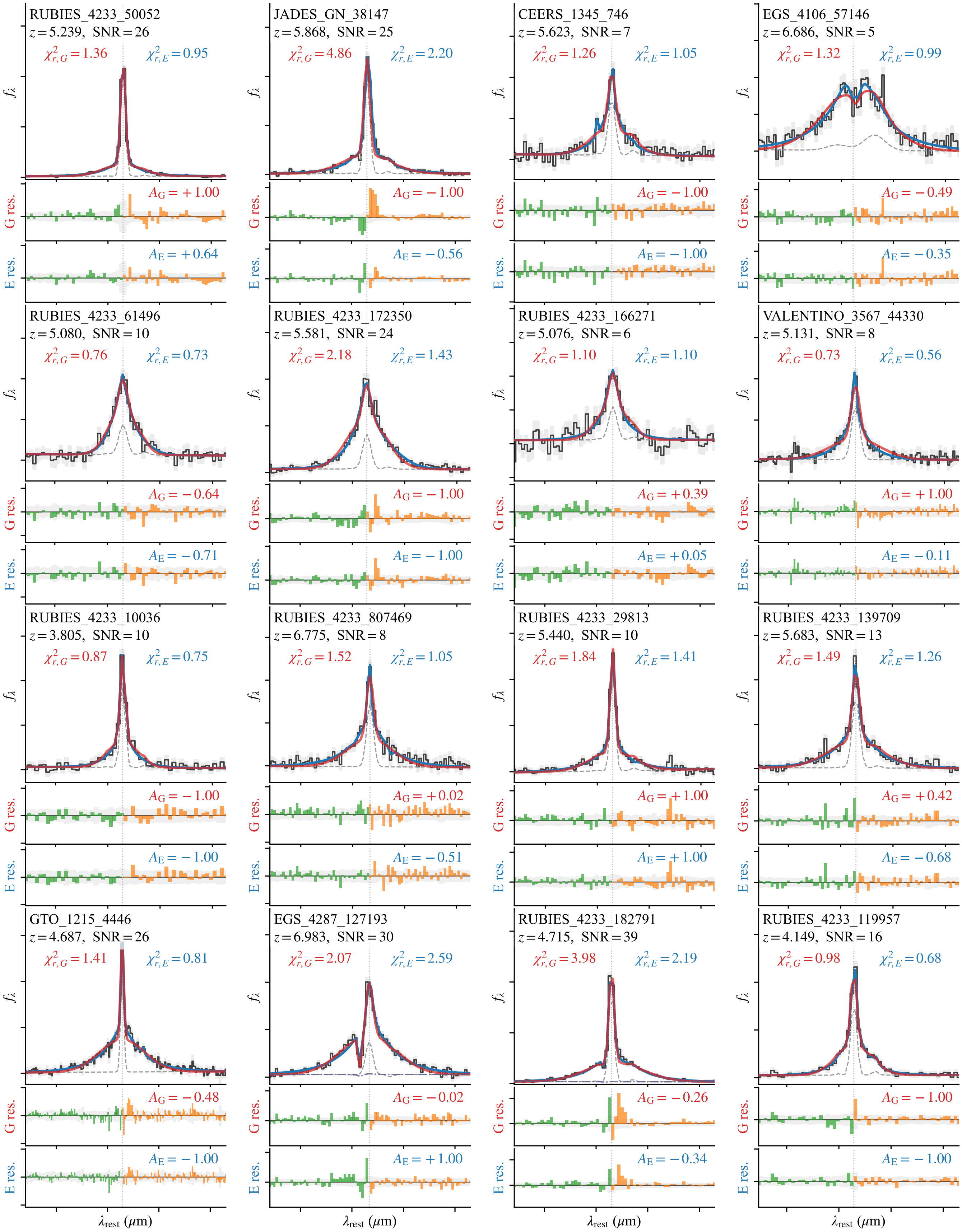

**Figure 7.** (Continued.)

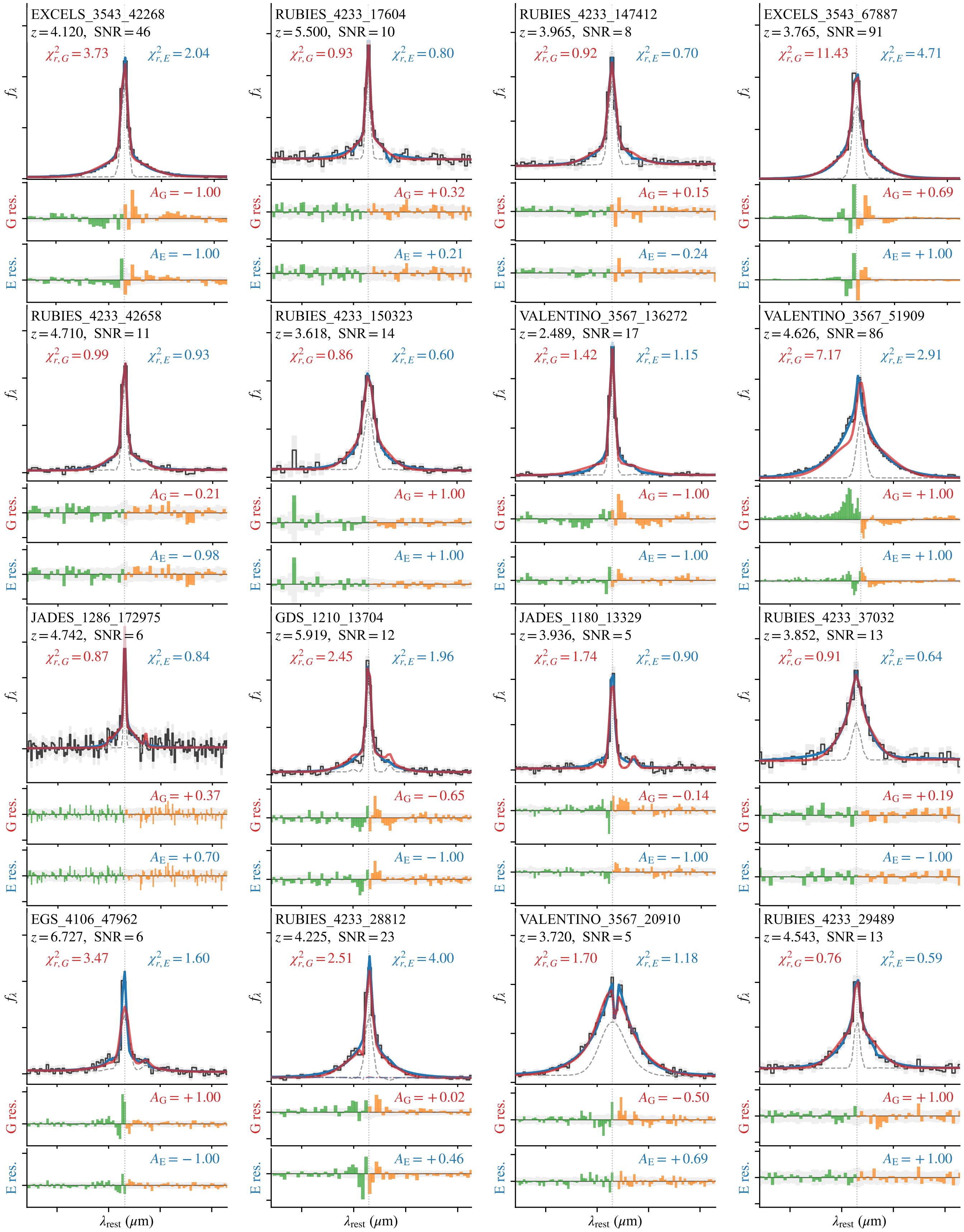

**Figure 7.** (Continued.)

To test this, we follow the methodology in V. Rusakov et al. (2026) to fit an exponential profile in contrast to the Gaussian profiles fitted previously (see Section 3.2) for all 48 known LRDs in our sample. Specifically, we choose to fit the H$\alpha$ + N II complex. Our results are shown in Figure 7. For both cases, we modulate the fits by a P Cygni profile or Gaussian absorption component when necessary, as described in V. Rusakov et al. (2026). We calculate the $\Delta\chi^2$ by

$$\Delta\chi^2 = \frac{\chi^2_{\rm Gau} - \chi^2_{\rm Exp}}{\chi^2_{\rm Gau}} * 100\%. \quad (11)$$

We find our objects to have a median of $\Delta\chi^2 = 24.9\%$, meaning the exponential fits are typically ~25% better than Gaussian fits. These results support the framework of V. Rusakov et al. (2026), who suggest the electron scattering in a quasi-spherical, dense ionized gas cocoon likely produces the broad lines in H$\alpha$. We also note that the Gaussian model can still fit well in 30/48 of our LRDs ($\chi^2 < 2$). Therefore, the two scenarios are not mutually exclusive under the current quality of data. This is likely because in some cases the SNR of the broad wings is not high enough to distinguish the two fitting methods.

In addition, we define the broad-line asymmetry via $A = (\Sigma_{\rm blue} - \Sigma_{\rm red})/(|\Sigma_{\rm blue}| + |\Sigma_{\rm red}|)$, where $\Sigma_{\rm blue}$ and $\Sigma_{\rm red}$ are the summed model residuals blueward and redward of the fitted line center, with $A \in [-1, +1]$ indicating blue- or red-wing excess, respectively. We find our objects to have a median $A_G = -0.016$ and $A_E = -0.013$ for Gaussian and exponential, respectively, suggesting the broad H$\alpha$ lines are close to symmetric on average.

The probable systematic preference for exponential profiles across our sample therefore provides independent support for the electron-scattering interpretation, which in turn implies that SMBH masses inferred from standard virial relations are overestimated by roughly 2 orders of magnitude at high redshift and bringing the LRD population into agreement with the local $M_{\rm BH} - M_*$ relation (J. E. Greene et al. 2026; V. Rusakov et al. 2026). In future papers, similar analysis of the larger samples of LRDs with higher SNR data could further help distinguish these two scenarios.

### 5.4. Caveats and Future Work

One limitation of our dataset is that the depths of the JWST observations vary across different surveys, which may lead to inhomogeneous BLAGN coverage. Future work would also benefit from additional high-resolution spectroscopy as the majority of our BLAGNs (~70%) are identified with medium-resolution data. Moreover, deeper spectroscopic observations are required to extend the accessible redshift range and to strengthen the robustness of broad-line identification. In particular, the SNR of the H$\beta$ line is often too low in early Universe sources ($z > 6$) to reliably detect the broad-line component, restricting the maximum redshift of our survey to $z \sim 7.2$. We find a tentative trend that the H$\beta$ SNR drops with decreasing exposure time. Considering our criteria of detecting broad lines from emission features with SNR > 5, we conclude that future observations need to have exposure time of $\sim 10^4$ s to securely identify BLAGNs solely from H$\beta$ (for high-redshift sources that do not have H$\alpha$ covered in NIRSpec).

## 6. Conclusion

In this work, we conducted a comprehensive and systematic search for BLAGNs using the full spectroscopic range of JWST NIRSpec. From its medium- and high-resolution spectra, we identify 227 BLAGNs spanning redshifts of 0.8–7.2. This expanded sample offers insights into the growth and evolution of SMBHs in the early Universe. Notably, the high-redshift AGNs of our sample are characterized by systematically lower black hole masses and luminosities compared to local AGNs, yet they exhibit higher accretion efficiencies, indicating a distinct evolutionary phase in black hole–galaxy coevolution at $z \gtrsim 4$, as shown in Figure 3. In addition, compared with low-$z$ BLAGNs from SDSS, we find that accretion rates are generally higher by a factor of around 5–10. However, at $z \gtrsim 2$, the accretion rates remain mostly constant (Figure 5). Together this suggests that high-$z$ SMBHs undergo a characteristic lifecycle that is largely independent of the epoch at which they form.

We cross-match our BLAGN sample with the literature and reidentify a population of 48 LRDs. We fit these LRDs with Gaussian and exponential components and consistently identify, in 42/48 cases, that LRDs are better fit by exponential profiles. It should therefore be considered that electron scattering may be the origin of the broad lines for LRDs, and thus the inferred black hole masses could be substantially lower, potentially by up to 2 orders of magnitude. Although LRDs in our dataset have a lower expected accretion rate for their mass than other objects we identify (Figure 3), this correction would increase their $\lambda_{\rm edd}$ and also bring LRDs back into closer agreement with the local $M_{\rm BH}$–$M_*$ relation. Our detection of a large population of faint, low-mass BLAGNs at high redshift helps bridge the observational gap between local SMBHs and the previously studied luminous quasars at high redshift, providing a more complete view of AGN demographics in the early Universe.

## Acknowledgments

C.B. acknowledges the Research Experiences in Astronomy at CIERA for High School Students (REACH) program, which initiated this project idea, and thanks J. Formato for the helpful discussions. X.X. acknowledges the fellowship funding from Center for Interdisciplinary Exploration and Research in Astrophysics (CIERA), Northwestern University.

The data products presented here were retrieved from the Dawn JWST Archive (DJA). DJA is an initiative of the Cosmic Dawn Center (DAWN), which is funded by the Danish National Research Foundation under grant DNRF140.

The authors have used ChatGPT (OpenAI, 2023; https://chat.openai.com/chat) to check the coding syntax and to check grammar.

## Appendix
## All Identified BLAGNs

In Figure 8 we show the spectra of all identified BLAGNs.

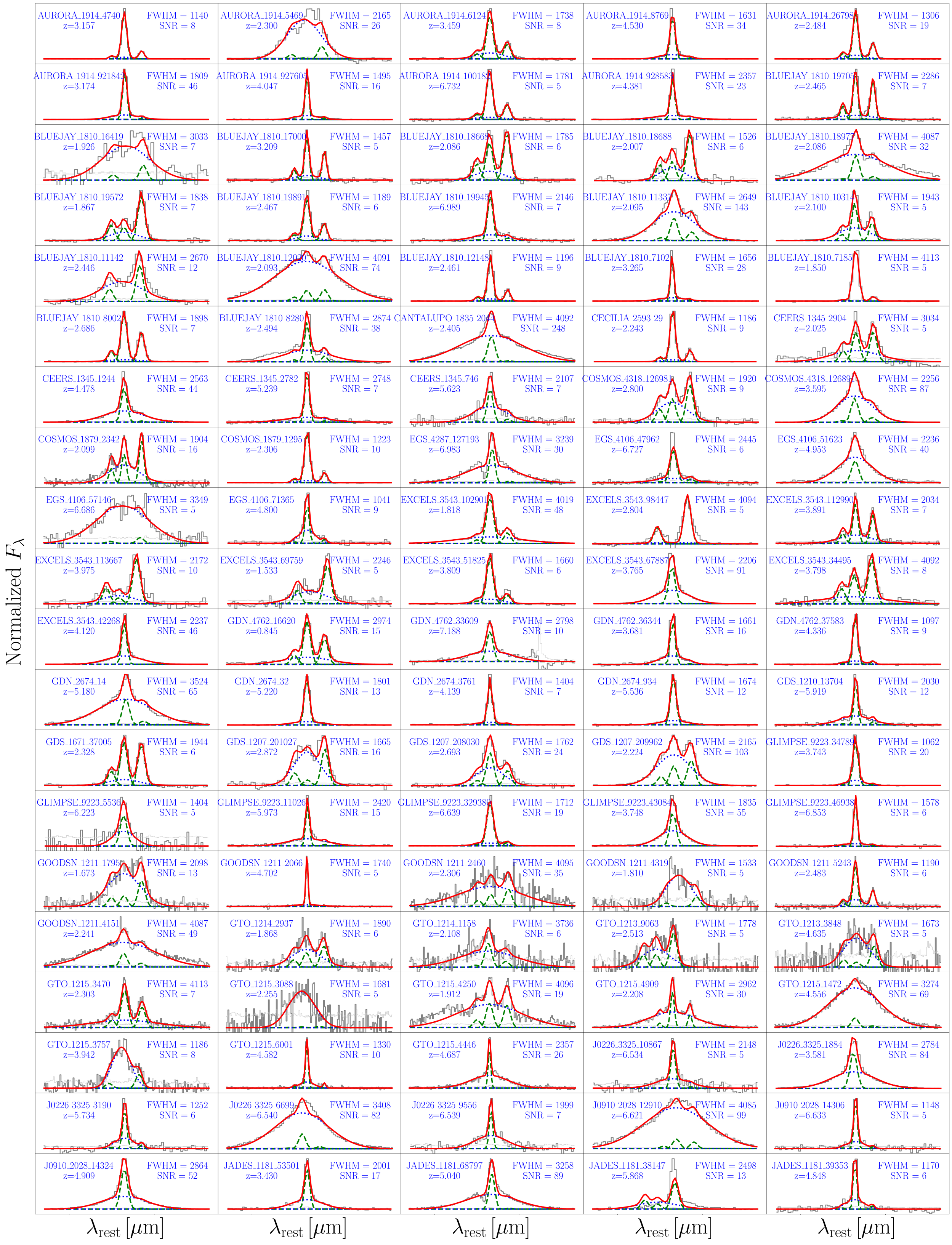

**Figure 8.** JWST NIRSpec spectra for identified BLAGNs. The data and error are shown by the black and gray lines, respectively. Green dashed and blue dotted lines represent the narrow and broad components, respectively. The sum of both components is shown in red. For each galaxy, we also overlay their redshift, H$\alpha$ broad component's FWHM (in km s$^{-1}$), and line signal to noise.

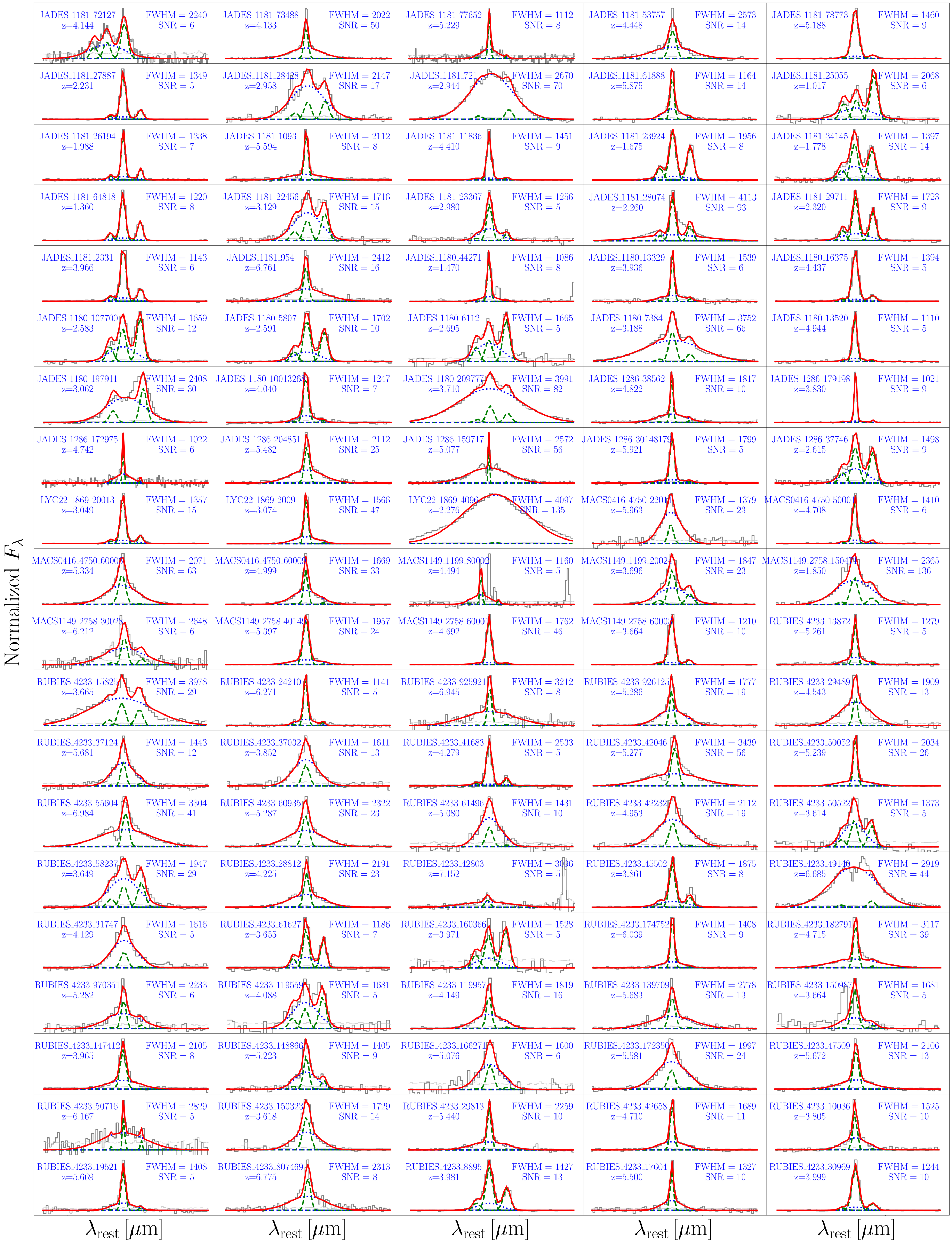

**Figure 8.** (Continued.)

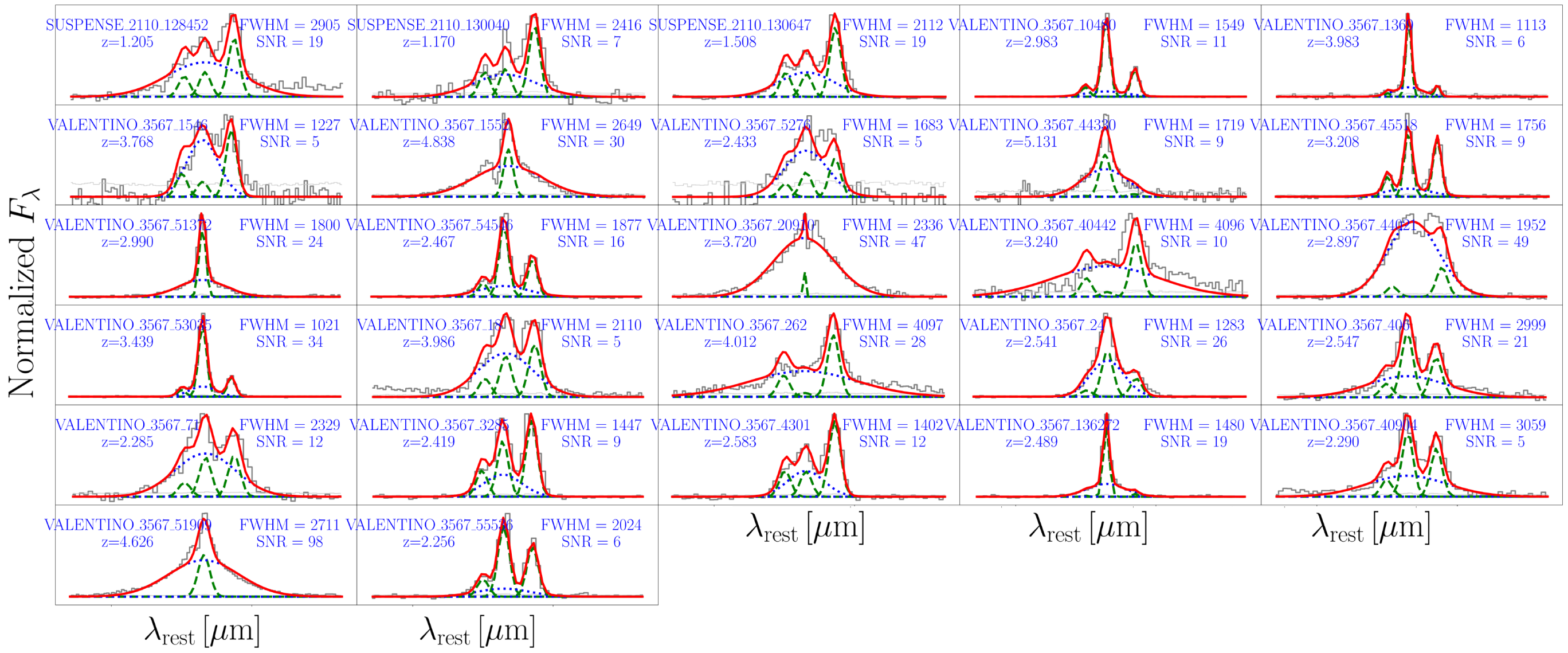

**Figure 8.** (Continued.)

## ORCID iDs

Caroline J. Baccus https://orcid.org/0009-0007-7847-3325
Xinfeng Xu https://orcid.org/0000-0002-9217-7051